\documentclass[preprintnumbers, showpacs, floatfix,preprintnumbers, letterpaper, superscriptaddress,nofootinbib,11pt]{revtex4}
\usepackage[utf8]{inputenc}
\usepackage{amsfonts}
\usepackage{amsmath}
\usepackage{amssymb,epsf}
\usepackage{latexsym}
\usepackage{graphicx,epsfig}
\usepackage{amssymb}
\usepackage{float}
\usepackage{subfigure}
\usepackage{epstopdf}
\usepackage[colorlinks=true,citecolor=blue,linkcolor=blue,urlcolor=black]{hyperref}
\usepackage{dcolumn}
\usepackage{psfrag}
\usepackage{wrapfig}
\usepackage{makeidx}
\usepackage{epsf}
\usepackage{color}

\begin{document}

\title{Mimetic gravity in $(2+1)$-dimensions}
\author{Ahmad Sheykhi\footnote{asheykhi@shirazu.ac.ir}}
\address{Physics Department and Biruni Observatory, Shiraz
University, Shiraz 71454, Iran\\
Research Institute for Astronomy and Astrophysics of
Maragha (RIAAM), P.O. Box 55134-441, Maragha, Iran\\
Max-Planck-Institute for Gravitational Physics
(Albert-Einstein-Institute), Am M$\ddot{u}$hlenberg 1, DE-14476
Potsdam-Golm, Germany}

\begin{abstract}
One of the most important achievements in general relativity has
been discovery of the $(2+1)$-dimensional black hole solutions of
Einstein gravity in anti-de Sitter (AdS) spacetime [Phys. Rev.
Lett. {\bf69}, 1849 (1992)]. In this paper, we construct, for the
first time, the $(2+1)$-dimensional solutions of mimetic theory of
gravity. These solutions may provide a powerful background to
investigate the physical properties of mimetic gravity and examine
its viability in lower spacetime dimensions. In particular, some
physical properties of stationary black hole solutions of this
theory in the presence of charge or angular momentum are
investigated.

\end{abstract}
 \maketitle

\section{Introduction }
Black holes are perhaps the most fascinating and amazing
astrophysical objects that have ever emerged as the solutions of
gravitational theories. These wonderful and enrich structures
provide a powerful background for exploring different branches of
physics including quantum gravity, thermodynamics, superconducting
phase transition, paramagnetism-ferromagnetism phase transition,
superfluids, condensed matter physics, spectroscopy, information
theory, holographic hypothesis, etc. This powerfully is mainly due
to the discovery of the well established correspondence between
gravity in $d$-dimensional anti-de Sitter (AdS) spacetime and the
conformal field theory (CFT) living on the boundary of
$(d-1)$-dimensional spacetime, known as $AdS_d/CFT_{d-1}$
correspondence or gauge/gravity duality. Recently, black holes
have also received a renewed attention after some significant
steps toward understanding the puzzle of information paradox
\cite{HPS1} as well as the shadow of the supermassive black holes
as the first results of $M87$ Event Horizon Telescope \cite{EHT}.
It was shown that there exist soft hairs, including soft gravitons
and/or soft photons, on the black hole horizon and the complete
information about their quantum state is stored on a holographic
plate at the future boundary of the horizon \cite{HPS1}. Black
hole entropy and microscopic structure near the horizon can be
understood through these soft hairs \cite{Afshar,HHPS,HPS2,Grum}.

One of the most significant achievements in black hole
physics has been the discovery of three dimensional solutions of
general relativity in AdS spacetime known as BTZ
(Banados-Teitelboim-Zanelli) black holes \cite{BTZ1}. In fact, the
$(2+1)$-dimensional solution of Einstein gravity provides a
simplified model to investigate and find some conceptual issues
such as black hole thermodynamics, quantum gravity, string and
gauge/gravity duality, holographic superconductors in the context
of the $AdS_{3}/CFT_{2}$ \cite{Car1,Ash,Sar,Wit1,Car2}. It has
been shown that the quasinormal modes in this spacetime coincide
with the poles of the correlation function in the dual CFT. This
gives quantitative evidence for $AdS_{3}/CFT_{2}$ \cite{Bir}.
Furthermore, BTZ black holes play a crucial role for improving our
perception of gravitational interaction in low dimensional
spacetimes \cite{Wit2}. In particular, it might shed some light on
the quantum gravity in three dimensions. Geometry of the spinning
$(2+1)$-dimensional black holes has been explored in \cite{BTZ2}.
It was shown that the surface $r=0$ is not a curvature singularity
but, rather, a singularity in the causal structure which is
everywhere constant and continuing beyond it would produce closed
timelike curves \cite{BTZ2}. The extension to include an electric
charge in addition to the mass and angular momentum have been
performed \cite{BTZ3,Fern}. The $(2+1)$-dimensional black holes
also provide a powerful background to explore one-dimensional
holographic superconductors \cite{Ren, Liu, Kord,mahya,Bina}. The
studies on $(2+1)$-dimensional solutions of gravitational field
equations have been extensively carried out in the literatures
(see e.g.
\cite{rin1,rin2,Clem,Car,Noj3D,Emp,Cad,Par,Hendi,She,Xu,Grum2}).

In this paper, we introduce the $(2+1)$-dimensional black hole
solutions in the context of mimetic gravity. The theory of mimetic
gravity was proposed a few years ago, as an alternative
description for the dark matter puzzle \cite{Mim1}. It was argued
that the mimetic field can encodes an extra longitudinal degree of
freedom to the gravitational field. Thus, the gravitational field
achieves, in addition to two transverse degrees of freedom, a
dynamical longitudinal degree of freedom which can play the role
of mimetic dark matter. Latter, it was shown that a modified
version of mimetic gravity can resolve the cosmological
singularities \cite{MimCos1} as well as the singularity in the
center of a black hole \cite{Cham3}. Besides, it has been
confirmed that the original setting of the mimetic theory predicts
that gravitational wave (GW) propagates at the speed of light,
ensuring agreement with the results of the event GW170817 and its
optical counterpart \cite{Sunny1,Sunny2}. It has also been shown
that this theory can explain the flat rotation curves of spiral
galaxies \cite{MimMOND,ShSa}. Mimetic theory of gravity has arisen
a lot of enthusiasm in the past few years both from the
cosmological viewpoint
\cite{MimCos,Dutta,Sep,Zh,Mat,Odin0,Odin1,Odin2,MimCos2,Gorj1,Gorj2,Gorj3,Gorj4,Fir,Leb,Cham2}
as well as black holes physics \cite{Der,Myr1,Myr2,OdinP,
Odinfr1,Odinfr2,Oik,Gorji2,
Nash1,Ch,Nash3,jibril,Yunlong1,Yunlong2,Shey}.

Till now, black hole solutions of mimetic gravity in
$(2+1)$-dimensions have not been explored. Our purpose here is to
construct static and stationary analytical black hole solutions of
mimetic gravity in three dimensional spacetime and investigate
their properties. These new solutions provide a set up,
for future investigations, to examine mimetic theory of gravity
and its physical consequences in lower spacetime dimensions and
examine the viability of this theory. In particular, it may be
useful for investigating one-dimensional holographic
superconductors in the context of mimetic gravity. We shall
consider several cases including whether or not there is a
coupling to the Maxwell field or whether or not there is an
angular momentum associated with the spacetime. We study the
effects of the mimetic field on the casual structure and physical
properties of the solutions and disclose that, in contrast to the
three dimensional solution of general relativity, in mimetic
gravity a curvature singularity emerges at $r=0$ even in the
absence of Maxwell field. This essential singularity might be due
to the extra longitudinal degree of freedom of the gravitational
field encoded by the mimetic field. Surprisingly, the curvature
singularity disappears by adding an angular momentum to the
spacetime.

This paper is organized as follows. Section \ref{Field} is devoted
to introducing the basic field equations of mimetic gravity in
$(2+1)$-dimensions. For simplicity we first ignore the coupling to
the Maxwell field in this section and construct three dimensional
black hole solutions. In Sec. \ref{Stat}, we take into account the
Maxwell field and explore charged mimetic black holes in three
dimensions. In Sec. \ref{Ang}, we add an angular momentum to the
back hole and investigate rotating $(2+1)$-dimensional solution of
mimetic gravity. We summarize our results in Sec. \ref{Sum}.

\section{Field equations and solutions} \label{Field}
We start with the following action
\begin{eqnarray}\label{Act}
&&S=\int{d^3x\sqrt{-{g}}\left(\mathcal{R}+{\lambda}(g^{\mu \nu
}\partial _{\mu} \phi
\partial _{\nu}\phi-\epsilon)+\frac{2}{l^2}-F_{\mu \nu }
F^{\mu \nu }\right)},
\end{eqnarray}
where $\mathcal{R}$ is the Ricci scalar, $\lambda$ is the Lagrange
multiplier, and $l$ is related to the cosmological constant by
$-\Lambda=l^{-2}$. $F_{\mu \nu }=\partial _{\mu }A_{\nu }-\partial
_{\nu }A_{\mu }$ is the electromagnetic field tensor and $A_{\mu
}$ is the gauge potential. In the above action, $\epsilon=\pm1$,
in which the positive and negative sign refer to, respectively,
spacelike and timelike nature of vector $\partial _{\mu} \phi$. We
adopt $(-,+,+)$ as our signature and set $8\pi G_N=1$ throughout
this work. The equations of motion can be derived from varying
action (\ref{Act}), yielding
\begin{eqnarray}\label{FE1}
{G}_{\mu\nu}&=& \lambda \partial _{\mu} \phi \partial _{\nu}
\phi+\frac{g_{\mu \nu }}{l^2}+T_{\mu \nu }
\end{eqnarray}
\begin{equation}\label{FE2}
\frac{1}{\sqrt{-g}}\partial _{\kappa}(\lambda \sqrt{-g} \partial
^{\kappa} \phi )=\nabla _{\kappa}(\lambda \partial ^{\kappa} \phi
)=0,
\end{equation}
\begin{equation}
\partial _{\mu }\left( \sqrt{-g} F^{\mu \nu }\right)
=0. \label{FE3}
\end{equation}%
\begin{equation}\label{cond}
g^{\mu \nu}\partial _{\mu} \phi \partial _{\nu} \phi=\epsilon.
\end{equation}
where \begin{equation}\label{Tem} T_{\mu \nu }=2 F_{\mu \gamma }
F_{ \nu }^{\ \gamma}-\frac{1}{2} g_{\mu \nu } F_{\alpha \beta }
F^{\alpha \beta },
\end{equation}
being the Maxwell energy-momentum tensor. Equation (\ref{cond})
restricts the evolution of the mimetic field $\phi$ and indicates
that the scalar field is not dynamical by itself, nevertheless it
makes the longitudinal degree of freedom of the gravitational
field dynamical \cite{Mim1}. It was argued \cite{Mim1} that if one
assume $g_{\mu\nu}=g_{\mu\nu} (\phi,\tilde{g}_{\mu \nu})$, in such
a way that  $g_{\mu \nu}= \epsilon (\tilde{g}^{\alpha \beta}
\partial _{\alpha} \phi
\partial _{\beta} \phi)\tilde{g}_{\mu \nu}$, then one recovers (\ref{cond}) immediately (see also \cite{Der,Fir}).
Tracing  Eq. (\ref{FE1}), combining with Eq. (\ref{cond}), yields
$\lambda=\epsilon\left(G-T-3/l^2\right),$ where $G$ and $T$ are,
the trace of the Einstein tensor and energy momentum tensor,
respectively. Substituting $\lambda$ in the field equations (\ref
{FE1}) and (\ref {FE2}), they transform to
\begin{eqnarray}\label{FEE1}
&&{G}_{\mu\nu}= \epsilon\left(G-T-\frac{3}{l^2}\right) \partial
_{\mu} \phi
\partial _{\nu} \phi+\frac{g_{\mu \nu }}{l^2}+T_{\mu \nu },\\
 &&\partial _{\kappa}\left[\sqrt{-g}\left(G-T-\frac{3}{l^2}\right) \partial ^{\kappa} \phi
\right]=0,\label{FEE2}
\end{eqnarray}
Our aim here is to derive static $(2+1)$-dimensional black hole
solutions of the above field equations. We assume the metric of
spacetime as
\begin{eqnarray}\label{metric1}
ds^{2} &=&-f(r) g^2(r) dt^{2}+\frac{dr^{2}}{f(r)}+ r^2 d\varphi
^{2},
\end{eqnarray}
where an additional degree of freedom is incorporated in the line
element through the metric function $g(r)$ which is expected to
reflect an extra degree of freedom of gravitation encoded by the
mimetic field $\phi$. With metric (\ref{metric1}), the constraint
Eq. (\ref{cond}) transforms to $f(r) \phi'^{2}=\epsilon$, which
has solution of the form
\begin{equation}\label{phi}
\phi(r)=\int{\frac{dr}{ \sqrt{|\epsilon f(r)|}}},
\end{equation}
where we have chosen the positive sign and set the integration
constant equal to zero, without loss of generality. Equation
(\ref{phi}) shows the explicit dependence of the mimetic field on
the metric function and reveals that it is not an independent
dynamical variable. Let us first consider the uncharged solution.
In this case the field equations (\ref{FEE1}) and (\ref{FEE2}),
have the following solutions
\begin{eqnarray}\label{fr1}
f(r)&=&-M+\frac{r^2}{l^2},\\
g(r)&=&1+\frac{b\ r}{\sqrt{|r^2-Ml^2|}}, \label{gr1}
\end{eqnarray}
where $b$ is the constant of integration which incorporates the
impact of the mimetic field into the solutions. In general $b$
could be either positive or negative. For $b=0$, our solutions
reduce to the $(2+1)$-dimensional BTZ black holes of Einstein
gravity \cite{BTZ1}. The horizon is located at $r_{+}=l\sqrt{M}$
where $f(r_{+})=0$, while $g(r_{+})$ diverges. However, $r=r_{+}$
is a coordinate singularity and both Kretschmann and Ricci scalars
have finite values at $r_{+}$. The sign of $|r^2-Ml^2|$ depends on
whether one considers interior solution ($r<r_{+}$) or exterior
solution ($r>r_{+}$). Expanding $g(r)$ for large $r$ leads to
\begin{equation}\label{gr1ex}
g(r)\approx 1+b+\frac{bMl^2}{2r^2}+ O\left(\frac{1}{r^4}\right).
\end{equation}
Therefore as $r\rightarrow \infty$, we have $g(r)\approx 1+b$,
which implies that the remnant of the mimetic field $\phi$
contributes to the metric function $g(r)$ through constant $b$.
Nevertheless the asymptotic behavior of the solutions is still AdS
since the constant $1+b$ can be absorbed by redefinition of the
time at the asymptotic region.

The ($tt$) component of the metric is given by
\begin{eqnarray}\label{gtt}
{-\textbf{g}_{tt}(r)}=f(r) g^2(r)=\frac{1}{l^2}\left[
\sqrt{|r^2-Ml^2|}+b \ r \right]^2,
\end{eqnarray}
where its expansion for large $r$ is given by
\begin{eqnarray}\label{gttexp}
{-\textbf{g}_{tt}(r)}\approx -(1+b)M+(1+b)^2 \frac{
r^2}{l^2}-\frac{bM^2 l^2}{4 r^2}+O\left(\frac{1}{r^4}\right).
\end{eqnarray}
This confirms that in large $r$ limit, we have
$-\textbf{g}_{tt}\neq\textbf{g}^{rr}$. On the other hand when
$r\rightarrow0$, we have $g(r)=1$ and
$-\textbf{g}_{tt}=\textbf{g}^{rr}=f(r)$. The infinite redshift
surface can be obtained by setting $\textbf{g}_{tt}(r)=0$, which
yields
\begin{eqnarray}\label{Irs}
r_{s_i}=\frac{r_{+}}{\sqrt{1\pm b^2}},
\end{eqnarray}
where $i=1,2$ and $-$ for $r_{s_1}>r_{+}$ and $+$ for
$r_{s_2}<r_{+}$. This means that we have 2 infinite redshift
surfaces and the black hole horizon is located between them,
$r_{s_2}<r_{+}<r_{s_1}$. Besides, $r_{s_1}$ exists provided
$b^2<1$.

We now calculate the scalar curvatures of the spacetime. It is a
matter of calculations to show that the Ricci scalar and the
Kretschmann invariant are given by
\begin{eqnarray}
&&R =\frac{2\left[b(Ml^2-3r^2)-3r \sqrt{|r^2-Ml^2|}\right]}{r l^2 \left(\sqrt{|r^2-Ml^2|}+b r\right)},  \label{Ric1} \\
&&R_{\mu \nu \rho \sigma }R^{\mu \nu \rho \sigma
}=\frac{4\Big{\{}3r^2(r^2-Ml^2)+2br(3r^2-Ml^2)
\sqrt{|r^2-Ml^2|}+b^2(3r^4-2Ml^2 r^2+M^2
l^4)\Big{\}}}{r^2 l^4 \left(\sqrt{|r^2-Ml^2|}+b r\right)^2},\nonumber \\
 \label{Kre1}
\end{eqnarray}
Thus, as $r\rightarrow0$, both Ricci and Kretschmann invariants
diverge, they are finite at $r\neq0$ and go to $R=-6/l^2$ and
$R_{\mu \nu \rho \sigma }R^{\mu \nu \rho \sigma }=12/l^4$ as
$r\rightarrow\infty$. Therefore, the spacetime has an essential
singularity at $r=0$. This is in contrast to the
$(2+1)$-dimensional black holes of Einstein gravity where $r=0$ is
not a curvature singularity but, rather, a singularity in the
causal structure \cite{BTZ2}. Indeed, for Einstein gravity where
$b=0$, the curvature invariants have constant values anywhere,
namely $R=-6/l^2$ and $R_{\mu \nu \rho \sigma }R^{\mu \nu \rho
\sigma }=12/l^4$, and the curvature singularity disappears.

In the next section, we shall consider the $(2+1)$-dimensional
charged black hole of mimetic gravity.
\section{Charged mimetic black holes in 3D} \label{Stat}
\begin{figure}[htp]
\begin{center}
\includegraphics[width=7cm]{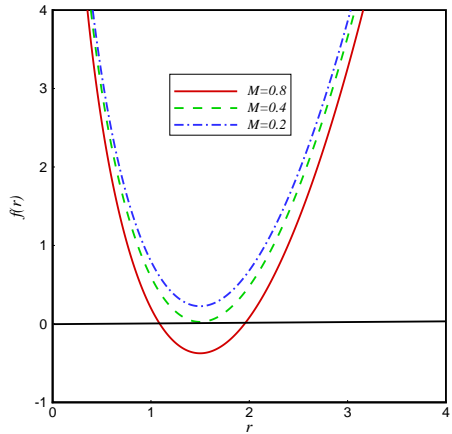}
\caption{The behavior of $f(r)$ for charged mimetic black holes in
3D and different $M$. Here we have taken $l=1$ and
$q=1.5$.}\label{Fig1}
\end{center}
\end{figure}

\begin{figure}[htp]
\begin{center}
\includegraphics[width=7cm]{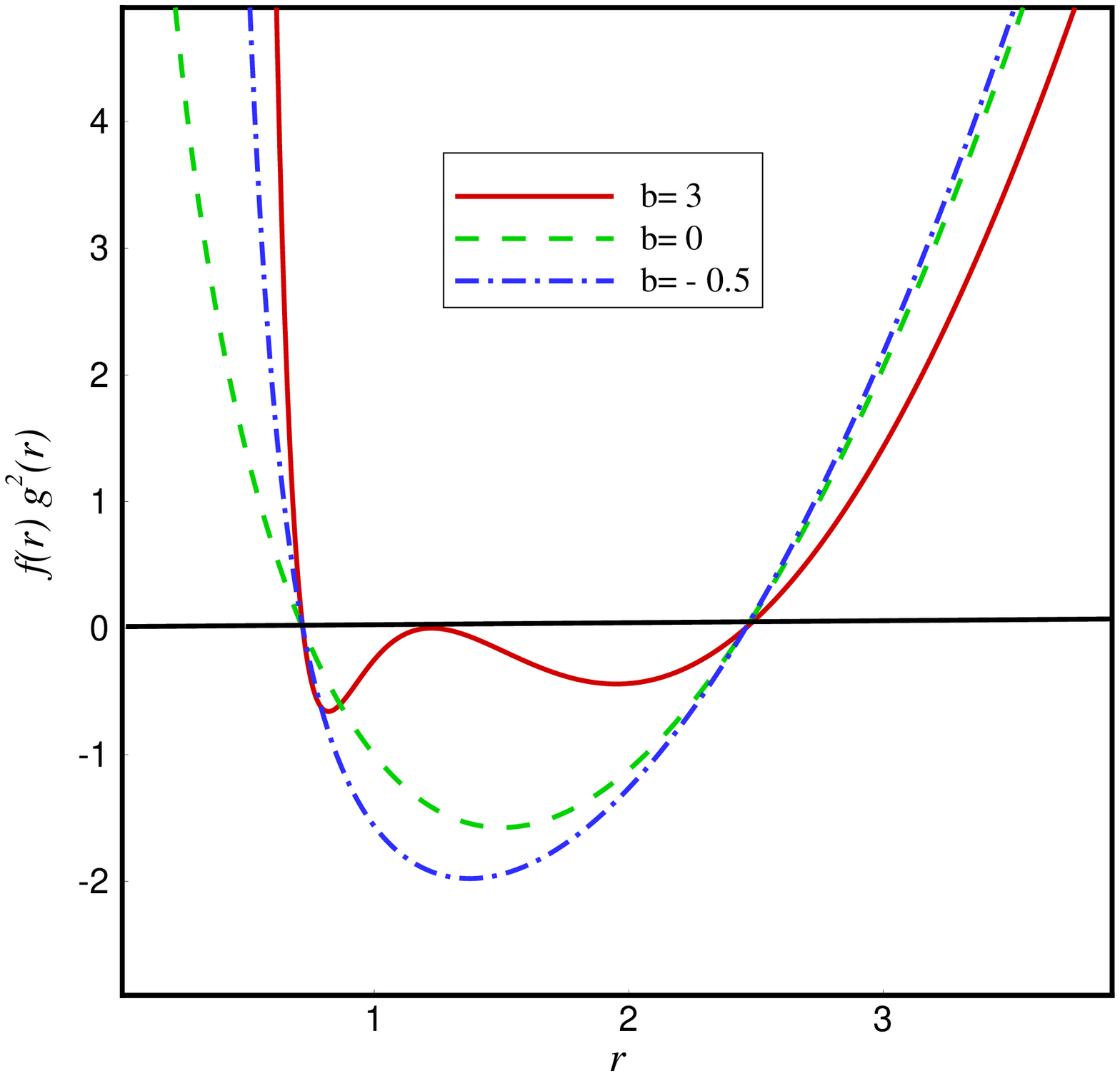}
\caption{The behavior of $-\textbf{g}_{tt}=f(r)g^2(r)$ for charged
mimetic black holes in 3D and different $b$. Here we have taken
$M=2$, $l=1$, $q=1.5$.}\label{Fig2}
\end{center}
\end{figure}

\begin{figure}[htp]
\begin{center}
\includegraphics[width=7cm]{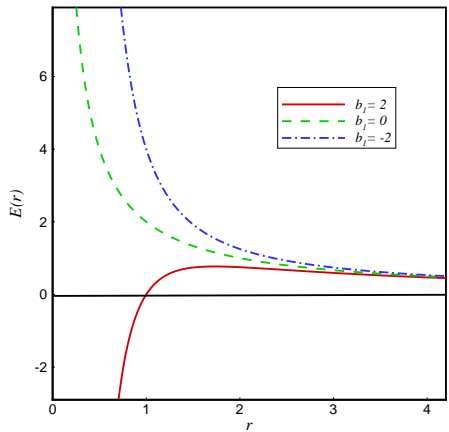}
\caption{The behavior of the electric field $F_{tr}=E(r)$ for 3D
charged mimetic black holes with $q=2$.}\label{Fig3}
\end{center}
\end{figure}

In the presence of coupling to the Maxwell field, the gauge
potential and the non-vanishing component of the electromagnetic
tensor become
\begin{equation}\label{At}
A_{\mu}= h(r) \delta^{0}_{\mu},  \  \  \  F_{tr}=h'(r),
\end{equation}
where prime indicates derivative with respect to $r$. Then it is
easy to show that the Maxwell equation (\ref{FE3}) transforms to
\begin{equation}\label{hr1}
rh''(r)g(r)-rh'(r)g'(r)+h'(r)g(r)=0,
\end{equation}
which has the following solution
\begin{equation}\label{Ftr1}
F_{tr}=h'(r)=\frac{q}{r}g(r),
\end{equation}
where $q$ is an integration constant which is indeed the electric
charge of the black hole. Substituting metric (\ref{metric1}),
condition (\ref{phi}), and the electric field (\ref{Ftr1}) into
Eq. (\ref{FEE1}), we find
\begin{eqnarray}\label{tt}
&&l^2 r f'- 2 r^2+2q^2 l^2=0, \\
&&l^2 g f'+3 l^2 r f' g'+2 l^2 r f g''+l^2 r g f''-4rg=0,\\
\label{rr} && 3l^2 r^2 f'g'+2l^2 r^2 f g''+l^2 r^2 g f''-2 r^2 g-2
q^2 l^2 g=0. \label{pp}
\end{eqnarray}
These equations have the following solutions
\begin{eqnarray}\label{fr2}
f(r)&=&-M-2q^2 \ln(r)+\frac{r^2}{l^2},\\
g(r)&=&1+b_1\int{\frac{dr}{\sqrt{|r^2-2 q^2 \ln (r)-Ml^2|^3}}},
\label{gr2}
\end{eqnarray}
where $b_1$ is again an integration constant which reflects the
imprint of the mimetic field on the spacetime. Clearly for
$b_1=0$, one recovers the $(2+1)$-dimensional charged black holes
of general relativity \cite{BTZ3}. One can easily check that these
solutions also satisfy Eq. (\ref{FEE2}) for the mimetic scalar
field. The horizons are given by the roots of $f(r_{h})=0$.
Depending on the parameters this equation may have at most two
real roots corresponding to inner and outer horizon of the black
hole. Of course, one may choose the parameters such that the
solutions also describe extremal black hole with one horizon, or a
naked singularity (see Fig. \ref{Fig1}). The integrand in
expression (\ref{gr2}) also diverges at $r_{h}$ indicating that we
encounter singularity. However, $r=r_{h}$ is only a coordinate
singularity and both Kretschmann and Ricci scalars have finite
values on the horizon. The integral in $g(r)$ function cannot be
analytically performed for arbitrary values of $r$, however, it is
instructive to study the large $r$ limit of $g(r)$. In this case,
one can write
\begin{eqnarray}\label{gr2ex}
g(r)&\approx&1+b_1\int{\frac{dr}{r^3}}\approx 1-\frac{b_1}{2 r^2},\\
-\textbf{g}_{tt}(r)&\approx & \left( -M-2q^2
\ln(r)+\frac{r^2}{l^2}\right)\left(1-\frac{b_1}{2 r^2}\right)^2 \\
\label{gtt2ex} &\approx & -M-\frac{b_1}{l^2}-2q^2
\ln(r)+\frac{r^2}{l^2}+\frac{2b_1q^2
ln(r)}{r^2}+\left(Mb_1+\frac{b_1^2}{4l^2}\right)\frac{1}{r^2}+O\left(\frac{1}{r^4}\right).\nonumber
\end{eqnarray}
Therefore, as $r\rightarrow \infty$ we have
$-\textbf{g}_{tt}\neq\textbf{g}^{rr}$, and the remnant of the
mimetic field contributes to the metric function $\textbf{g}_{tt}$
through constant $b_1$. It is also interesting to take a close
look on the roots of $\textbf{g}_{tt}=0$ which define the infinite
redshift surfaces. Since there is no any physical reason to avoid
negative $b_1$, thus one can choose either $b_1>0$ or $b_1<0$. For
$b_1\leq0$, the infinite redshift surfaces $r_{s_i}$ coincide with
the horizons $r_h=r_{\pm}$, namely the roots of $f(r_h)=0$.
However, for $b_1>0$, the function
 $\textbf{g}_{tt}$ admits an additional root $r_{s_3}=\sqrt{b_1/2}$ (see Fig.
 \ref{Fig2}). Note that $r_{s_3}$ can be either larger, equal or smaller
than $r_{-}$ depending on the values of $b_1$. It is easy to show
that, for all values of $b_1$, both Ricci and Kretschmann scalars
diverge at $r=0$, they are finite at $r\neq0$ and as $r\rightarrow
\infty$ they tend to constant values, $R=-6/l^2$ and $R_{\mu \nu
\rho \sigma }R^{\mu \nu \rho \sigma }=12/l^4$, similar to 3D
solutions of Einstein gravity. This analysis confirms that there
is a curvature singularity at $r=0$, regardless of the value of
$b_1$. The behaviour of the electric field for the large $r$ is
also given by
\begin{equation}\label{Ftr1ex}
F_{tr}\approx\frac{q}{r}\left(1-\frac{b_1}{2 r^2}\right).
\end{equation}
We have also plotted the behavior of the electric field $F_{tr}$
in Fig. \ref{Fig3}. This figure shows that the electric field
diverges for small $r$ and goes to zero in large $r$ limit. For
$b_1>0$, the electric field has a maximum at finite $r$ which can
be easily seen from expression (\ref{Ftr1ex}). Expression
(\ref{Ftr1ex}) also shows that, compared to the case of Einstein
gravity, the electric field of three dimensional charged mimetic
black hole get modified due to the presence of mimetic field.


\newpage
\section{Rotating mimetic black holes in 3D} \label{Ang}
Now we consider spinning solutions of mimetic gravity in
$(2+1)$-dimensions. We take the spinning metric as
\begin{eqnarray}\label{metric2}
ds^{2} &=&-f(r) g^2(r) dt^{2}+\frac{dr^{2}}{f(r)}+ r^2 (J
N(r)dt+d\varphi ^{2})^2,
\end{eqnarray}
where functions $f(r)$, $g(r)$ and $N(r)$ are determined by
solving the field equations. Clearly, in the limiting case where
the mimetic gravity reduces to Einstein gravity, one expects to
have $g(r)=1$ and $N(r)=-1/2r^2$ \cite{BTZ1}. The $(t\phi)$
component of the field equations (\ref{FEE1}) yields
\begin{eqnarray}\label{tphi}
r g(r) N''(r)+3g(r)N'(r)-r N'(r)g'(r)=0,
\end{eqnarray}
which implies $g(r)=r^3 N'(r)$. Combining with $(tt)$ component of
the field equation we arrive at
\begin{eqnarray}\label{tphi}
l^2 J^2 +2l^2r^3 f'(r)-4r^4=0,
\end{eqnarray}
with the following solution
\begin{eqnarray}\label{frJ1}
f(r)=-M+\frac{r^2}{l^2}+\frac{J^2}{4r^2}.
\end{eqnarray}
Finally, the $(rr)$ component of the field equations (\ref{FEE1})
can be solved to give
\begin{eqnarray}\label{NrJ}
N(r)&=&-\frac{1}{2
r^2}+\frac{b_0}{r^2}\sqrt{|l^2J^2-4Ml^2r^2+4r^4|},\\
g(r)&=&r^3 N'(r)=1+\frac{2b_0
Ml^2\left(2r^2-\frac{J^2}{M}\right)}{\sqrt{|l^2J^2-4Ml^2r^2+4r^4|}}.
\label{grJ}
\end{eqnarray}
One can easily show that these solutions fully satisfy the field
equations (\ref{FEE1}) and (\ref{FEE2}). In the limiting case
where $b_0=0$, the obtained solutions reduce to the
$(2+1)$-dimensional rotating black hole solutions
 of general relativity \cite{BTZ1}. When $J=0$,
they restore the solutions (\ref{fr1}) and (\ref{gr1}) provided we
define $b=2b_0 Ml^2$. It is a matter of calculations to check that
$f(r)$ vanishes for two values of $r$ given by
\begin{eqnarray}\label {rhJ}
r_{\pm}= l\sqrt{\frac{M}{2}\left(1\pm \sqrt{1-\frac{J^2}{M^2
l^2}}\right)},
\end{eqnarray}
where $r_{+}$ is the black hole horizon. In order for the solution
to describe a black hole, one must have
\begin{eqnarray}\label {MJ}
M>0, \  \   \   \   |J|\leq M l.
\end{eqnarray}
In the extremal case where $|J|=Ml$, the two roots coincide and we
have $r_{+}=r_{-}=l\sqrt{M/2}$.

The surface of infinite redshift can be given by finding the
positive real roots of the following equation
\begin{eqnarray}\label {gtt3}
\textbf{g}_{tt}(r)=4b_0^2 r^2(J^2-M^2l^2)+M-2Mb_0
\sqrt{l^2J^2-4Ml^2r^2+4r^4}-\frac{r^2}{l^2}=0,
\end{eqnarray}
with following solution
\begin{eqnarray}\label {rsJ}
r_{s}= l{\sqrt{\frac{M(1+ 2b_0lJ)}{ 1+ 4b_0lJ+4b_0^2l^2
(J^2-M^2l^2) }}}.
\end{eqnarray}
Therefore, the solution  admits a surface of infinite redshift,
similar to the $(2+1)$-dimensional rotating black hole of general
relativity \cite{BTZ2} which has an infinite redshift surface
located at $r_s=l \sqrt{M}$. In the limiting case where $b_0=0$
the infinite redshift surface coincides with one of Einstein
gravity, while for the extremal case ($|J|=Ml$) the infinite
redshift surface becomes
\begin{eqnarray}\label {rsext}
r^{ext}_{s}= l{\sqrt{\frac{M(1+ 2b_0 M l^2)}{ 1+ 4b_0Ml^2}}}.
\end{eqnarray}
In general the location of $r^{ext}_{s}$ depends on the value of
$b_0Ml^2$ and it is easy to check that it is located out of
horizon, namely $r^{ext}_s>r_{+}$ where $r_{+}=l\sqrt{M/2}$ is the
horizon radius of extremal case.

Next we study the scalar invariants. It is easy to check that as
$r\rightarrow0$, the Ricci and Kretschmann invariants behave as
\begin{eqnarray}
&&\lim_{r\rightarrow0}R =\frac{2[2b_0lJ^2+4b_0l^3M^2-3J]}{l^2 J(1-2b_0 l J)},  \label{Ricci3} \\
&&\lim_{r\rightarrow 0}R_{\mu \nu \rho \sigma }R^{\mu \nu \rho
\sigma }=\frac{4\left[ 3J^2(1+4b_0^2l^2J^2)-16b_0^2 l^4 M^2
(J^2-M^2l^2)-4b_0 l J (J^2+2M^2 l^2)\right]}{l^4 J^2(1-2b_0 l
J)^2},\nonumber \\\label{Kret3}
\end{eqnarray}
which have finite values unless for the static case ($J=0$), where
both of them diverge at $r=0$. This is a very interesting result
which reveals that in $(2+1)$-dimensional mimetic gravity, adding
the angular momentum $J$ to the spacetime removes the curvature
singularity at $r=0$. Indeed for $J=0$, the Ricci and Kretschmann
invariants reduce to (\ref{Ric1}) and (\ref{Kre1}) with
replacement $b=2b_0 Ml^2$  and both of them diverge at $r=0$. On
the other hand, in the asymptotic region where $r\rightarrow
\infty$ we have still $R=-6/l^2$ and $R_{\mu \nu \rho \sigma
}R^{\mu \nu \rho \sigma }=12/l^4$ for rotating solutions
($J\neq0$), which confirms that the solutions are asymptotically
AdS similar to $(2+1)$-dimensional rotating solutions of general
relativity.
\begin{figure}[htp]
\begin{center}
\includegraphics[width=7cm]{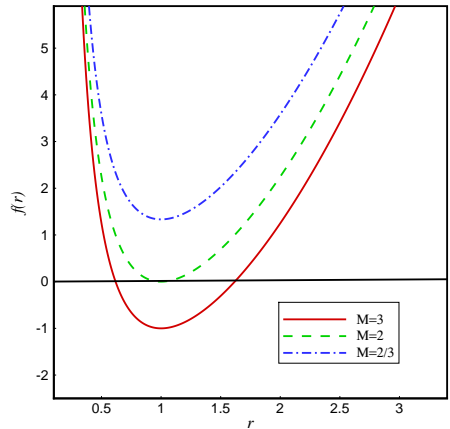}
\caption{The behavior of $f(r)$ for rotating black hole in 3D with
different $M$, where we have taken $l=1$ and $J=2$.}\label{Fig4}
\end{center}
\end{figure}
\begin{figure}[htp]
\begin{center}
\includegraphics[width=7cm]{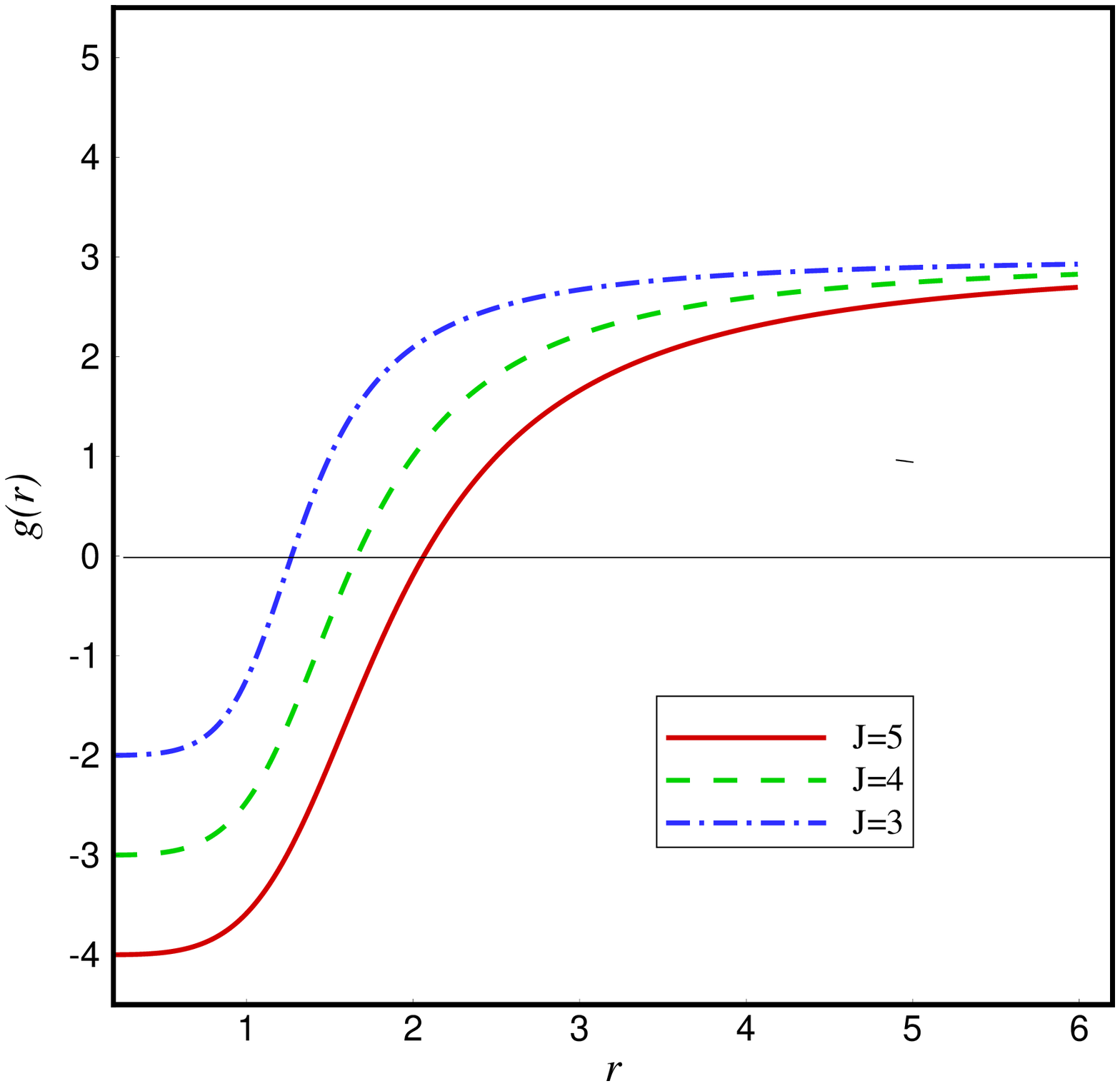}
\caption{The behavior of $g(r)$ for rotating black hole in 3D with
different $J$, where we have taken $l=1$, $b_0=0.5$ and
$M=2$.}\label{Fig5}
\end{center}
\end{figure}
\begin{figure}[htp]
\begin{center}
\includegraphics[width=7cm]{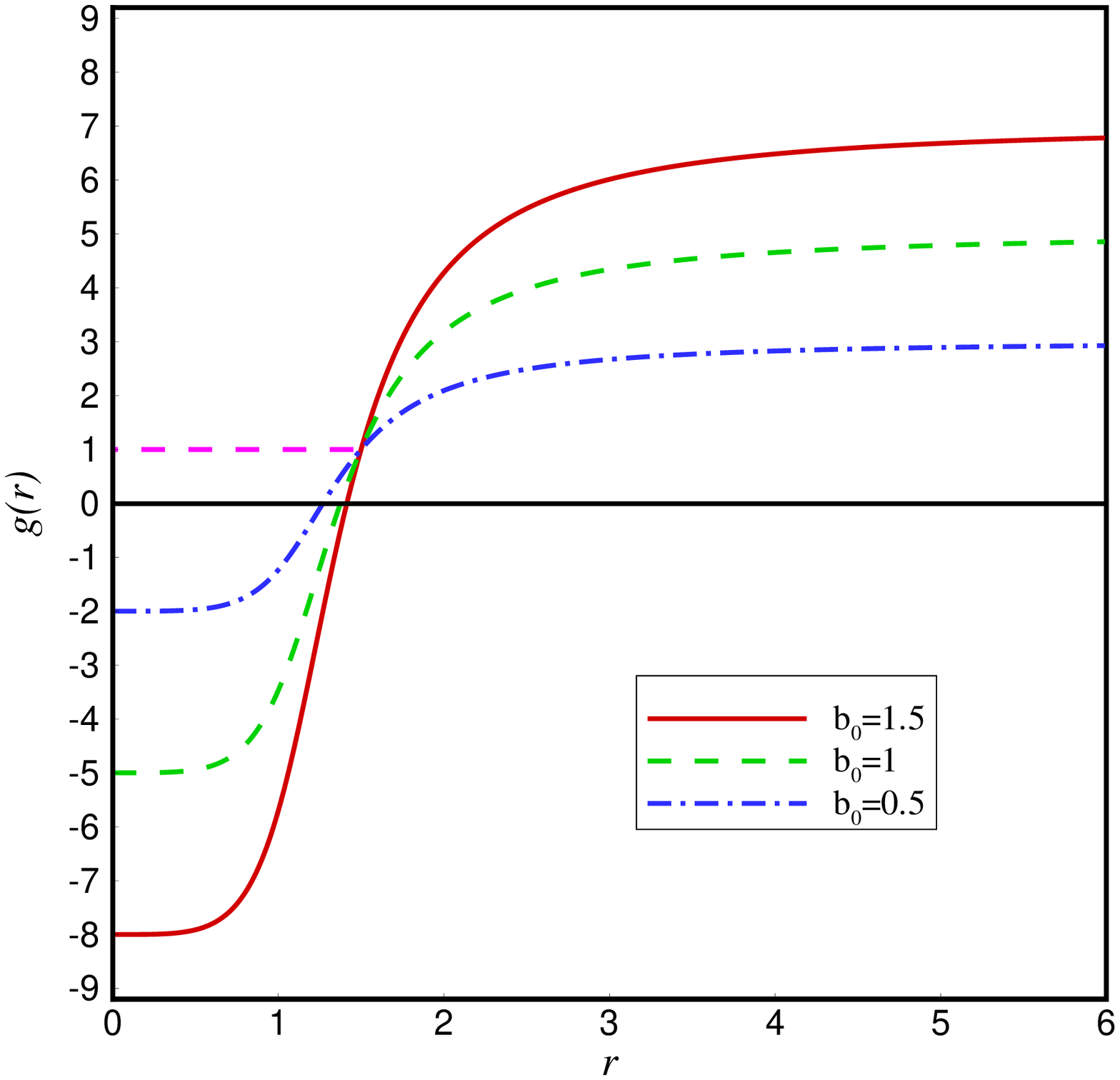}
\caption{The behavior of $g(r)$ for rotating black hole in 3D with
different $b_0$, where we have taken $l=1$, $J=3$ and
$M=2$.}\label{Fig6}
\end{center}
\end{figure}
\begin{figure}[htp]
\begin{center}
\includegraphics[width=7cm]{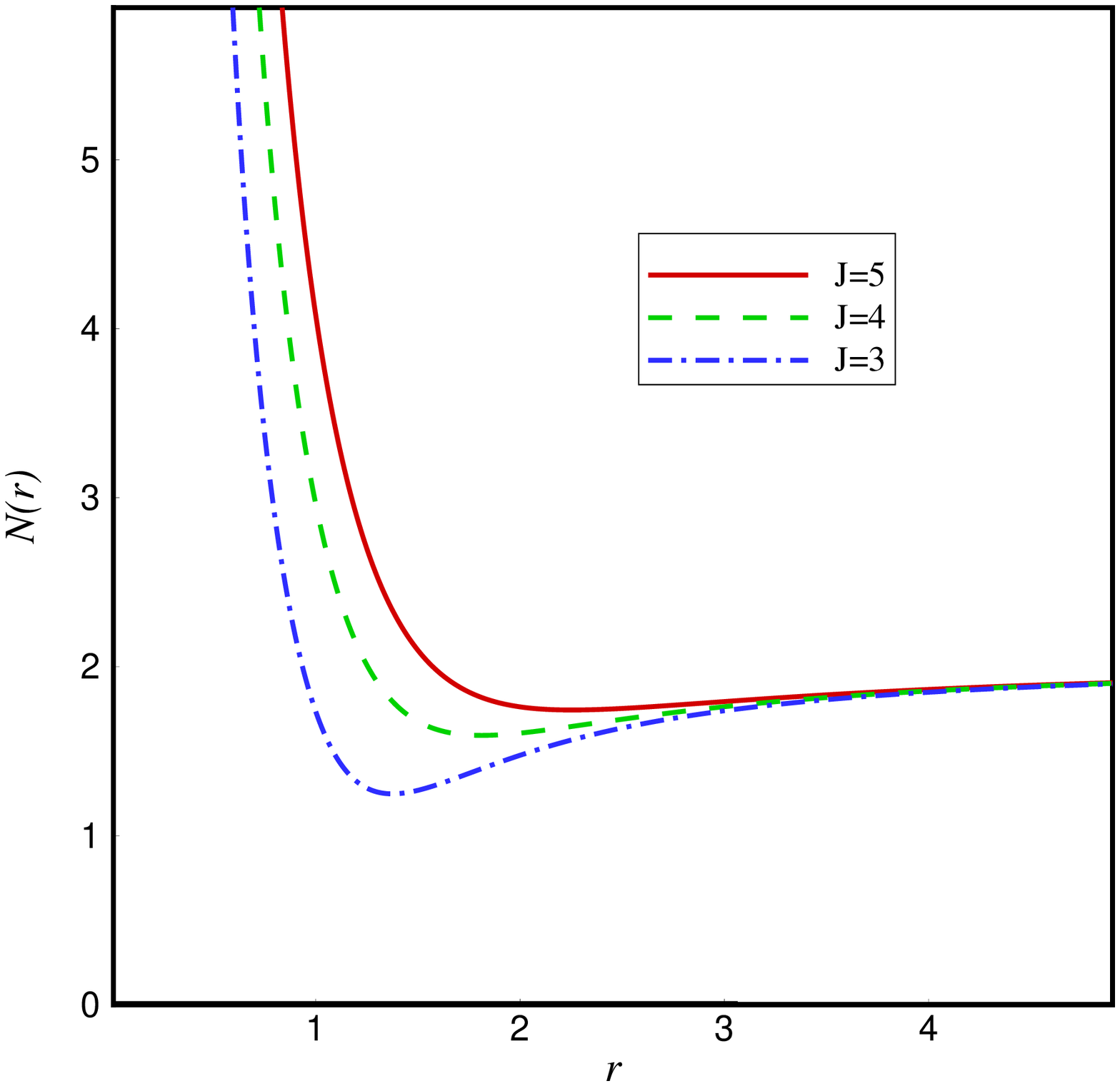}
\caption{The behavior of $N(r)$ for rotating black hole in 3D with
different $J$, where we have taken $l=b_0=1$ and
$M=2$.}\label{Fig7}
\end{center}
\end{figure}
\begin{figure}[htp]
\begin{center}
\includegraphics[width=7cm]{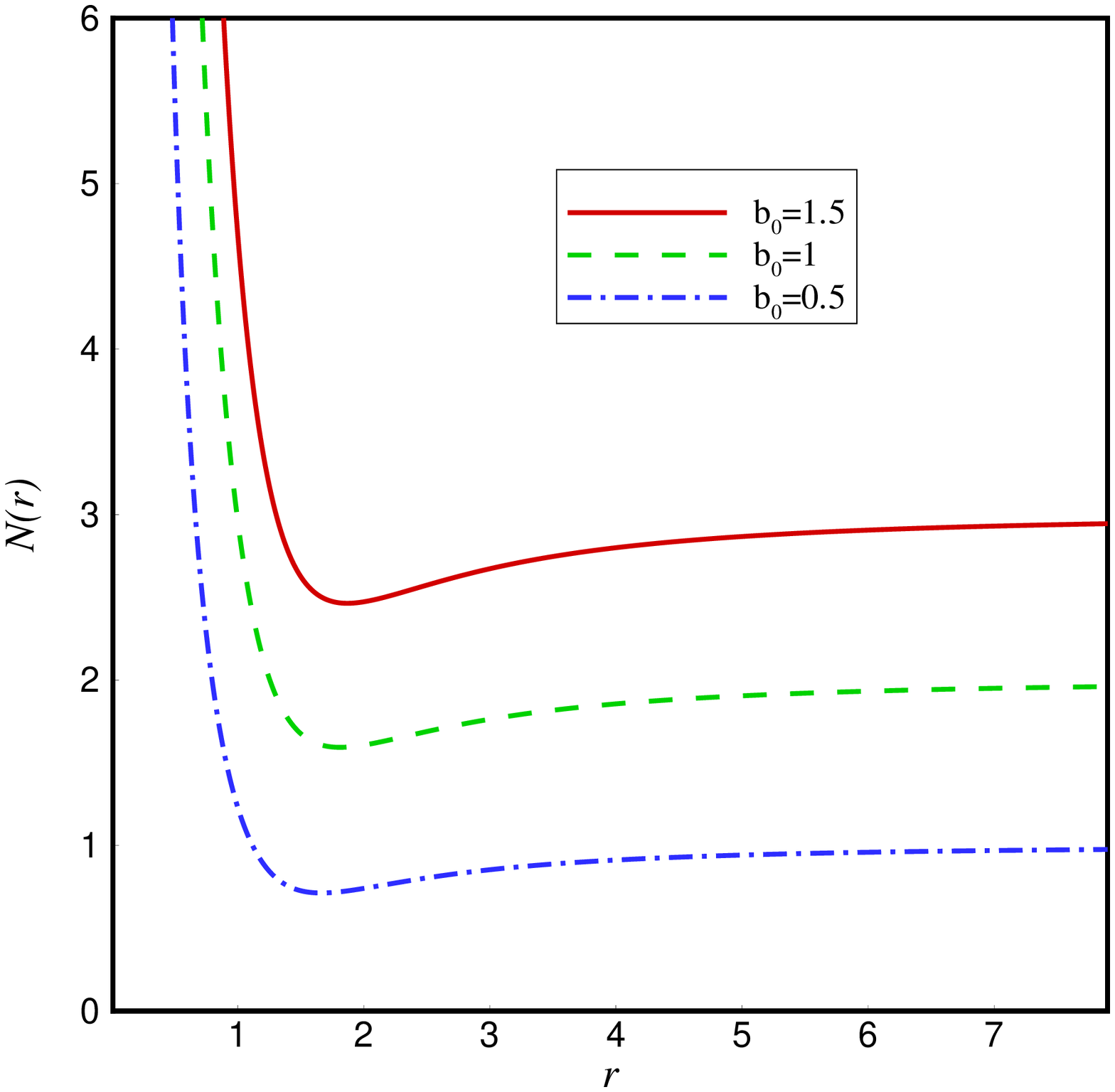}
\caption{The behavior of $N(r)$ for rotating black hole in 3D with
different $b_0$, where we have taken $l=1$, $J=4$ and
$M=2$.}\label{Fig8}
\end{center}
\end{figure}
\newpage
Let us remind that for $b_0=0$ the curvature invariants have
finite values at the origin for both static and rotating
solutions, namely $R=-6/l^2$ and $R_{\mu \nu \rho \sigma }R^{\mu
\nu \rho \sigma }=12/l^4$ \cite{BTZ2}. In mimetic gravity,
however, we observed that while for static solutions they diverge
at $r=0$, but for rotating solution they have finite values at
$r=0$. In other words, imprint of mimetic field to the spacetime
structure in mimetic theory of gravity makes the spacetime
singular at the origin $r=0$, while the combination of angular
momentum $J$ and mimetic field remove the curvature singularity at
the origin.

The behavior of the metric functions $f(r)$, $g(r)$ and $N(r)$ for
rotating mimetic black holes in three dimensions are plotted in
Figs. \ref{Fig4}-\ref{Fig8}. From Fig. \ref{Fig4}, we see that
depending on the metric parameters, our solutions can represent
black hole with one horizon, two horizon or naked singularity.
Figs. \ref{Fig5} and \ref{Fig6} reveal that the metric function
$g(r)$ tends to a constant value far from the black hole. The
value of this constant depends on $b_0$, $M$ and $l$ but
independent of $J$. The intersection of all curves in Fig.
\ref{Fig6} is the point in which $g(r)=1$ which occurs at
$r=J/\sqrt{2M}$ independent of $b_0$. We have also depicted the
behavior of $N(r)$ in Figs. \ref{Fig7} and \ref{Fig8} where it can
be seen that $N(r)\rightarrow \infty$ as $r\rightarrow0$, while
for large values of $r$ we have $N(r)\rightarrow 2b_0$.

\section{Summary}\label{Sum}
To sum up, using a scalar mimetic field for isolation the
conformal degree of freedom of the gravitational field in a
covariant way, it has been demonstrated that the scalar field can
encode an extra dynamical longitudinal degree of freedom to the
gravitation field which can play the role of mimetic dark matter
even in the absence of particle dark matter \cite{Mim1}. In this
paper, we have focused on mimetic gravity in $(2+1)$-dimensional
spacetime and constructed various static and spinning black hole
solutions of this theory. In contrast to the three dimensional
solutions of Einstein gravity which has only causal singularity
and scalar invariants are constant everywhere, in mimetic gravity
the spacetime admits a curvature singularity. We confirmed that
both Ricci and Kretschmann curvatures of three dimensional mimetic
black holes diverge at $r=0$ even in the absence of Maxwell field.
Interestingly, when the angular momentum is added to the
spacetime, the singularity at $r=0$ disappears and the scalar
invariants become constant.


\acknowledgments{I thank Shiraz University Research Council. I am
grateful to Max-Planck-Institute for Gravitational Physics (AEI),
where this work written and completed, for hospitality. This work
has been financially supported by the Research Institute for
Astronomy \& Astrophysics of Maragha (RIAAM), Iran.}


\end{document}